\documentclass[aps,pra,twocolumn,superscriptaddress,showkeys,showpacs]{revtex4-1}

\usepackage{graphicx}
\usepackage{amssymb,amsmath,amsfonts,latexsym,dsfont}
\usepackage[usenames,dvipsnames]{color} 
\definecolor{LinkColor}{rgb}{0,0,.5}
\usepackage[colorlinks=true , citecolor=blue,urlcolor=blue]{hyperref}
\usepackage{etoolbox} 

\usepackage[capitalize]{cleveref}
\usepackage{CJK}

\usepackage{algorithm}
\usepackage{algorithmic}

\renewcommand{\emph}{\textit}

\renewcommand{\emph}{\textit}

\newcommand{\R}{\mathbb{R}}

\graphicspath{{pics/}{}}

\makeatletter
\appto{\appendix}{%
  \@ifstar{\def\theequation@prefix{A.}}%
          {}%
}
\makeatother

\begin{document}
\begin{CJK*}{UTF8}{}
\title{Novel Trotter formulas for digital quantum simulation}

\author{Yi-Xiang Liu \CJKfamily{gbsn}(刘仪襄)}
\affiliation{
Research Laboratory of Electronics, Massachusetts Institute of Technology, Cambridge, Massachusetts 02139, USA.
} 
\affiliation{Department of Nuclear Science and Engineering, Massachusetts Institute of Technology, Cambridge, Massachusetts 02139, USA.
}
\author{Jordan Hines}
\affiliation{
Research Laboratory of Electronics, Massachusetts Institute of Technology, Cambridge, Massachusetts 02139, USA.
}
\affiliation{Department of Physics,  Massachusetts Institute of Technology, Cambridge, Massachusetts 02139, USA.
}

\author{Zhi Li \CJKfamily{gbsn}(李智)}
\affiliation{
Department of Physics and Astronomy, University of Pittsburgh, and Pittsburgh Quantum Institute, Pittsburgh, Pennsylvania 15260, United States
}

\author{Ashok Ajoy}
\affiliation{
Department of Chemistry, University of California Berkeley, and Materials Science Division Lawrence Berkeley National Laboratory, Berkeley, California 94720, USA.
}

\author{Paola Cappellaro}
\thanks{pcappell@mit.edu}
\affiliation{
Research Laboratory of Electronics, Massachusetts Institute of Technology, Cambridge, Massachusetts 02139, USA.
}\affiliation{Department of Nuclear Science and Engineering, Massachusetts Institute of Technology, Cambridge, Massachusetts 02139, USA.
}

\begin{abstract}
Quantum simulation promises to address many challenges in fields ranging from quantum chemistry to material science, and high-energy physics, and could be implemented in noisy intermediate-scale quantum devices. 
A challenge in building good digital quantum simulators is the fidelity of the engineered dynamics given a finite set of elementary operations.
Here we present a framework for optimizing the order of operations based on a geometric picture, thus abstracting from the operation details and achieving computational efficiency.  
Based on this geometric framework, we provide two alternative second-order Trotter expansions, one with optimal fidelity at a short time scale, and the second robust at a long time scale. 
Thanks to the improved fidelity at different time scale, the two  expansions we introduce can form the basis for experimental-constrained digital quantum simulation.   
\end{abstract}

\maketitle
\end{CJK*}

Simulation has been at the core of quantum information processing right from its inception, starting from Feynman's vision of simulating physics using a quantum system~\cite{Feynman82}. 
Quantum simulators  are posed to be one of the first quantum devices to show task-specific quantum supremacy~\cite{Preskill12x}. 
Quantum simulation  has great potential impact on quantum chemistry~\cite{OMalley16,Aspuru-Guzik05,Hempel18}, material science~\cite{Babbush18},  condensed matter~\cite{Lanyon11,Salathe15,Langford17,Wei18}, and high-energy physics~\cite{Martinez16,Zohar16}. 
The most flexible strategy to achieve quantum simulation is via digital quantum simulation~\cite{Lloyd96},  where a target time-evolution operator is represented by a sequence of elementary quantum gates, usually involving one or two qubits.  
The strategy of approximating the continuous evolution with discrete gates is also known as Trotter expansion~\cite{Trotter1959}. 
First- and second-order Trotter expansions with simple alternating patterns are most commonly used, although there has been a recent interest  also in randomized Trotter expansions~\cite{Childs18x,Campbell19}.

Finding a sequence of elementary operations that improves the fidelity to a desired simulated Hamiltonian   is crucial to practical implementations of digital quantum simulation.
Experimental platforms include  superconducting qubits~\cite{Lamata18,Langford17,Salathe15,Malley16},  trapped ions~\cite{Arrazola16,Lanyon11}, atomic systems~\cite{Weimer10}, and spin systems~\cite{Cai13,Wei18,Ajoy13l}.

While the fidelity of digital quantum simulation can be increased by increasing the number of Trotter steps,  on all experimental platforms, the smallest Trotter step will eventually hit some practical limitations. In many cases, it is desirable to engineer the simulated evolution from a finite set of elementary  operations (gates). For example, in fault-tolerant quantum computation, the  circuit is built up from a set of universal gates that can be implemented fault tolerantly~\cite{Nielsen00b,Jones_2012}. 
In Hamiltonian simulation, the rotation angle under a specific Hamiltonian can not be made arbitrarily small due to experimental constraints. For example, in a recent implementation of digital quantum simulation with trapped ions~\cite{Lanyon11}, the smallest flip angle of the unitary evolution block was $Jt=\pi/16$, and similar numbers can be obtained in other experimental platforms~\cite{Barends15}. Then, even the second-order Trotter sequence might not yield good enough fidelity, while higher-order expansions are usually hard  to implement  experimentally  due to the complicated, sometimes negative~\cite{Suzuki91} and even imaginary~\cite{Janke1992}, coefficients.

In this paper, we focus on finding second-order Trotter expansions that yield a better fidelity than the conventional second-order Trotter (2T) expansion \cite{Suzuki1986,Suzuki1990}, given the set of implementable gates. Finding the optimal sequence that yields the best fidelity requires integer optimization over the product of large operators describing  the exact form of each gate, which can be computationally expensive.  
Here we present a geometric framework to optimize the  sequence without calculating the quantum mechanical propagators. Using this picture, we present two expansions: one minimizes the third-order Trotter error, and we call it 2-Optimal (2O); the second one has better performance at a longer time scale, and we name it 2-Diagonal (2D). Both methods have better fidelity than the 2T expansion  and even 3rd-order Trotter when the time step is larger.


\begin{figure}
  \centering
    \includegraphics[width=0.5\textwidth]{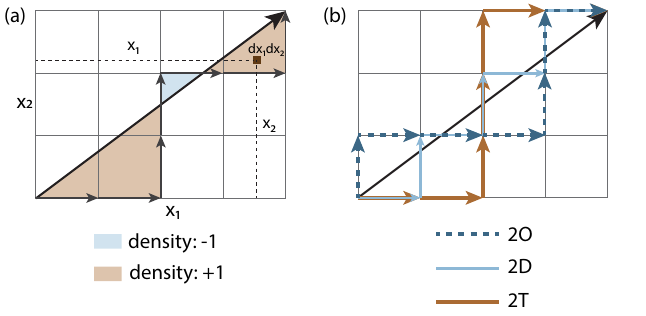}
  \caption{\textbf{Geometric method to calculate the Trotter error.} 
{(a)} Geometric picture for Trotter expansion of $e^{4A+3B}$ (vector (4,3)) by products of $e^A$'s and $e^B$'s. Unit vector (1,0) represents $e^A$ and unit vector (0,1) represents $e^B$. Any path starting from (0,0) that ends at (4,3) is accurate to first order. The second-order error is the  total area enclosed by the diagonal and the path, where any region below(above) the diagonal has an area density $+1$($-1$). The 3rd-order error is given by the moments, where, e.g., the infinitesimal moment about $x_2$-axis is the area of the dark brown square times the distance to $x_2$-axis.  
{(b)} Examples of three different second-order paths. The brown line is the conventional 2T expansion ($e^{2A}e^{3B}e^{2A}$).  The dark blue dashed path has a global minimum third-order error (2O, $e^{B}e^{3A}e^{2B}e^A$). The light blue path has a minimal total distance from each node to the diagonal (2D, $e^Ae^Be^Ae^Be^A$), represented by the sum of absolute areas. All three paths enclose zero total area.}
\label{fig:geometric}
\end{figure}

\begin{figure*}[t]
  \centering
	\includegraphics[width=1\textwidth]{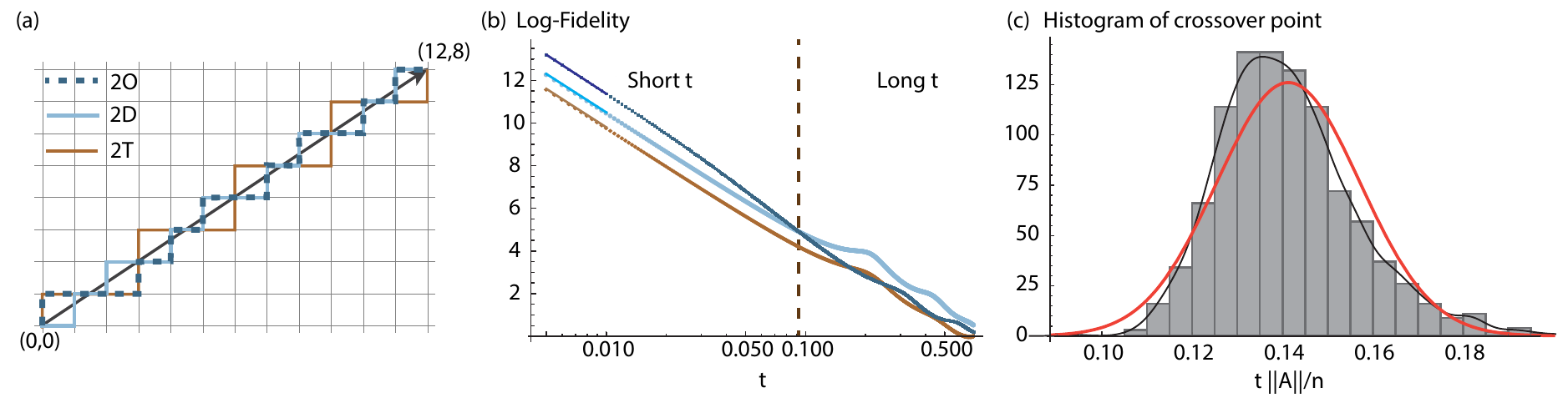}
  \caption{\textbf{Fidelity comparison for the three Trotter sequences.}  
(a) Geometric illustration of three different second-order approximations to $e^{12A+8B}$. In (a) and (b) dark blue lines are for the 2O path,  light blue for 2D, and  brown for 2T.  (b) Numerical evaluation of log-fidelity.  Here $A=- i H_1 t/n$, $B= - i H_2 t/n$, where $H_1 = \frac{1}{2}(\sigma_z^1 + \sigma_z^2$), $H_2 = \sigma_x^1 \sigma_x^2$, and $n=1$.  At short $t$, 2O shows the best fidelity since it has the smallest third-order error. At very small $t$ we fit the simulated data to $F_l=-\log_{10}(1-F) = -a \log_{10}t +b$,  and the slopes of 2O, 2D, and 2T are 6.07, 5.99, and 5.99 respectively, close to the expected result, 6, for second-order expansions. 2D starts to outperform 2O around $t=0.13 n/\|A\|$, where the $\|A\|$ is the Frobenius norm of $A$.   (c) Histogram of crossover point (where 2O and 2D intersect) for 1,000 pairs of random $4\times4$ Hermitian matrices $H_{1,2}$. The black solid curve is the smoothed histogram and the orange curve is a Gaussian fit to the histogram. The mean and variance of the Gaussian distribution are 0.14 and 0.02.  } 
\label{fig:numerics}
\end{figure*}

\paragraph*{Geometric method to calculate the Trotter error --}
We consider the problem of engineering a target operator $e^{-it\sum_L w_L H_L}$, which encapsulates a general quantum simulation task, using implementable gates $e^{-i H_k t/n}$. 
For illustration purpose let's first consider $L=2$ and  the target is $e^{pA+qB}$, where $A=-i H_1 t$ and $B=-i H_2 t$. Here we  assume  $p$ and $q$ to be integers and that the implementable gates are $e^{A/n}$ and $e^{B/n}$ (where $n>0$ is also an integer). Under the constraints mentioned above, we can generally write any expansion as
\begin{equation}\label{eq:QI}	
e^{pa_1A/n}e^{qb_1B/n}\dots e^{pa_MA/n}e^{qb_MB/n} \approx e^{p A + qB},
\end{equation}
where $\sum_{k\!=\!1}^M\!a_k\!=\!\!\sum_{k=1}^M\!b_k\!=\!n$ and $pa_k$, $qb_k$ are integers $\forall k$. 
The unitary fidelity $F_U$ of expansion (\ref{eq:QI}) up to $\mathcal{O}(t^3)$ is 
\begin{align}
&F_U=e^{-(p A + qB)}(e^{pa_1A/n}e^{qb_1B/n}\dots e^{pa_MA/n} e^{qb_MB/n})= \nonumber \\
&\openone\! +\!\mathcal{E}_2[A,B] \! +\!\mathcal{E}_{3,A}[A,\![A,B]] \! +\!\mathcal{E}_{3,B}[B,\![A,B]]\! +\!\mathcal{O}(t^4)
\label{eq:fid}
\end{align}
where $ \mathcal{E}_2$ is of  order  $t^2$,  and $\mathcal{E}_{3,A},~\mathcal{E}_{3,B}$ are of order  $t^3$. 
For example, the first-order Trotter expansion $(e^{pA/n}e^{qB/n})^n$ gives $F_U= \openone +\frac{1}{2n}pq[A,B]+\mathcal{O}(t^3)$. If $p$ (or $q$) is even, we can engineer the 2T expansion, 
\begin{align}\label{eq:trotter2}
&(e^{pA/(2n)}e^{qB/n}e^{pA/(2n)})^n,\qquad \textrm{with}   \\
&F_U\!=\! \openone -\frac{1}{24n^2} pq(p[A,[A,B]]+2q[B,[A,B])+\mathcal{O}(t^4)\nonumber
\end{align}
 to achieve a better approximation (a similar expression holds for $q$ even). 

In general, finding the coefficients $\{pa_k,qb_k\}$ is a hard problem since it requires an integer optimization search performed by calculating $F_U$ based on the exact form of the Hamiltonians $H_{1,2}$,  which might be prohibitive when considering large quantum systems. 
Here we present a geometric method to approach this algebraic problem, which enables us to optimize the ordering without considering the exact forms of the Hamiltonians.

For any expansion  in Eq.~\eqref{eq:QI}, we can accurately calculate the Trotter error for all possible paths up to at least the 3rd-order using a simple grid picture, as shown in Fig.~\ref{fig:geometric}(a). 
The ideal operator $e^{pA+qB}$ can be represented on the grid as a vector $(np,nq)$. One elementary gate is represented by a unit vector: $e^{A/n}$ is the unit vector (1,0) and $e^{B/n}$ is the unit vector (0,1). 
Starting from the origin $(0,0)$, at  each step we can move right  (evolving under $e^{A/n}$)  or up  (evolving under $e^{B/n}$). A given ordering is then represented as a directed path on the grid. Fig.~\ref{fig:geometric}(b) shows three different orderings for $p=4, q=3$ and $n=1$. 
This grid representation can be regarded as a ``projection'' of the Trotterized process on the unitary group. More rigorously, the Trotterized unitary operators live in a fiber bundle, for which the grid is the base space (see Appendix.~\ref{sec:geometry} for details).
On this grid picture, any path that ends at $(np,nq)$ is accurate to first order. The second-order error of the path is given by (see Appendix.~\ref{sec:geometry} for derivation)
\begin{align}
\frac{1}{2}[A,B](\int x_1dx_2-x_2dx_1)=[A,B]\iint dx_1dx_2,	
\label{eq:integral2}
\end{align}
where the coefficient of  $[A,B]$ can be interpreted as the total area defined by the path and the diagonal: $\iint dx_1dx_2$, where the  areas below (above) the diagonal have density $+1$ ($-1$). Then, to cancel the second-order error, the area should sum up to zero.
The third-order error is (see Appendix.~\ref{sec:geometry} for derivation)
\begin{align}\label{eq:integral}
	&\frac{1}{3}[A,[A,B]]\int x_1(x_1dx_2-x_2dx_1)\nonumber \\
&\qquad+\frac{1}{3}[B,[A,B]]\int x_2(x_1dx_2-x_2dx_1)\\
&=[A,[A,B]]\iint x_1dx_1dx_2
+[B,[A,B]]\iint x_2dx_1dx_2.\nonumber
\end{align}

   In the third-order error, the coefficient of $[A,[A,B]]$($[B,[A,B]]$) is the total moment about the $x_2$-axis($x_1$-axis) defined by the path [see  Fig.~\ref{fig:geometric}(a)].   If all steps are restricted to be directional, that is, either along $(0,1)$ or $(1,0)$, the third-order error is always non-zero~\cite{Suzuki91}.  
In principle, a third-order expansion, which has zero total moments about both axes, can also be represented by this picture, by allowing walking left (-1,0) ($e^{-A/n}$) and down (0,-1) ($e^{-B/n}$). This indeed leads, e.g., to the known 3rd-order Trotter expansion known as Ruth's formula~\cite{Ruth83}. Given experimental constraints that typically do not allow simple inversion of the Hamiltonian arrow of time, we do not consider  this scenario.

Using this geometric framework, we can now discuss two second-order sequences optimizing over different cost functions. The first sequence (2O) has a global minimum third-order error. The other sequence (2D) stays as close to the diagonal as possible and minimizes the distance from each node on the path to the diagonal. 2D has a better performance at a longer time scale, even if it is not  optimal in terms of the third-order error. The algorithms used to find the two sequences have different computational complexities.

\paragraph*{2-Optimal Sequence --}
In order to minimize the third-order error,  we start by assigning each edge on the grid a triplet of weights given by Eq.~\eqref{eq:integral2} and the first line of Eq.~\eqref{eq:integral}.   Using dynamic programming, we can find a path that puts the second-order error to zero and minimizes the third-order error. 
The main idea is that at each step, we keep track of the accumulated weights for each node, and finally select the path corresponding to the smallest absolute value weight. The details of the algorithm can be found in Appendix.~\ref{sec:DynamicProgramming}. 
In general, the  optimization should be done over all Trotter steps, thus on a grid $pn\times qn$. When the expansions become deep and optimizing over all Trotter steps is too expensive to calculate, one can optimize within a coprime $(p,q)$ and repeat, or mirror symmetrize the smallest unit~\footnote{When both $p$ and $q$ are odd, it is not possible to cancel the second-order error and one needs to perform the optimization on  a $(2p,2q)$  unit grid.}.

\paragraph*{2-Diagonal Sequence --}
This algorithm is not based on the exact expression of the Trotter error, but rather on the  intuition that being as close to the diagonal as possible should give a good approximation. There are different metrics to quantify ``being close"; 2D algorithm  minimizes the  distance from each node to the diagonal, thus minimizing the total distance, which can be shown to minimize the total unsigned area enclosed by the diagonal and the path as well. 
What is interesting is that the global distance optimal path can be found in a computationally efficient way by a  greedy optimization (see  Appendix~\ref{sec:D} for the proof): at each step, we choose the move that ends closer to the diagonal. 
It can be shown that when $p,~q$ are mutually prime, and when at least one of them is even, 2D always finds a  unique path, and the  2D path guarantees the second-order error to be zero. When both $p$ and $q$ are odd, 2D finds two opposite-order paths with  opposite second-order error. We can then symmetrize the sequence to cancel the second-order error. 
That is, 2D can be made into a second-order expansion by optimizing and symmetrizing  on the smallest unit grid ($2(p+q)$ steps), and the whole sequence is generated by repeating the smallest unit. 

For some $(p,q)$s, 2D and 2O share the same ordering, but in general, 2D  does not have a global minimum third-order error. Still,  we can prove that 2D  still has smaller 3rd-order error compared with the 2T expansion. The details of the proof can be found in  Appendix~\ref{sec:appendix}.

Both 2O and 2D can be easily generalized to higher dimension (see Fig.~\ref{fig:3D}(b) in Appendix.~\ref{sec:3D}), that is, to the scenario where we want to combine a larger number of propagators, $e^{A/n},e^{B/n},e^{C/n}$, etc. Fig.~\ref{fig:geometric}(b) shows 2O, 2D, and 2T paths for $p=4,~q=3$ and $n=1$ on the grid picture. Fig.~\ref{fig:numerics}(a) illustrates paths for $p=12,~q=8$.
Especially when higher dimensions are involved, the simpler greedy optimization of 2D is beneficial to reduce the computational complexity.

\paragraph*{Numerical evaluations --}
To demonstrate and compare the advantage of 2D and 2O sequences, we numerically evaluate their fidelity at short and long time scale. 
Though the dynamic programming time and space complexities are polynomial in the grid dimension, the computation for 2O quickly becomes both time-consuming, and space-consuming as $p$ and $q$ become larger. Here we only compare expansions up to $p+q=20$.

The metric we use to quantify the performance is the average fidelity of a quantum gate~\cite{Nielsen02}, which reads 
\begin{align}
\label{eq:fidelity}
F&=|\text{Tr}(U_1^{\dag}U_2)	|/\text{Tr}(U_1^{\dag}U_1),
\end{align}
where $U_1$ is the ideal operator and $U_2$ is the approximated one. To further highlight differences in fidelity, we plot the log-fidelity $F_l=-\log_{10}(1-F)$.   We also considered  $\|U_1-U_2\|$, where $\|\cdot\|$ is Frobenius norm, as a metric of error, and the results are qualitatively the same. 

The orderings of the three  second-order Trotter expansions for $p\!=\!12,~q\!=\!8$ are shown in Fig.~\ref{fig:numerics}(a). Fig.~\ref{fig:numerics}(b) shows numerical  results for a two-spin transverse-field Ising Hamiltonian, $H=12 H_1+8H_2$, where  $H_1 = \frac{1}{2}(\sigma_z^1 + \sigma_z^2$) and $H_2 = \sigma_x^1 \sigma_x^2$. Then $A=- i H_1t$, $B= - i H_2t$.
For different  $(p,q)$  pairs (here with $p+q=20$) the results look qualitatively the same. We  also verified that the same qualitative results apply for different Hamiltonians, such as single-particle operators,  larger transverse-field Ising models with $N=2-10$ spins, and high-dimension random Hermitian matrices. 

We can first verify that all three expansions are second-order expansions with respect to $t$ by fitting the data to $F_l=-a \log_{10}t+b$ and extract the slope $a$ at small $t$ (see the solid lines in Fig.~\ref{fig:numerics}(b)). From the Taylor expansion we expect $a=6$ and the fitted values for 2O, 2D, 2T are 6.07, 5.99, and 5.99 respectively.   In the short $t$ regime, 2O shows the best fidelity,  as expected since it minimizes the third-order error. 2D also provides a higher fidelity than the conventional second-order  Trotter (2T) expansion for all $t$.  At some $t$, 2D even starts to outperform 2O,  the crossover point being around $t\|A\|/n =0.14$. 
Fig.~\ref{fig:numerics}(c) shows a histogram of crossover point positions over 1,000 pairs of random hermitian matrices $H_{1,2}$. The center of the Gaussian depends highly on specific $(p,q)$, but it is always on the order of 0.1. 
For typical experimental values, such as $Jt=\pi/16$~\cite{Lanyon11}, and $U\Delta t/2=5/16$~\cite{Barends15},  2D  is then the best sequence. 

Conversely, these results imply that 2D does not need as small a Trotter step to reach the same fidelity. Then, the number of elementary gates needed to reach a given fidelity is smaller, a practical advantage for digital simulation~\cite{Campbell19}, especially when fault-tolerant gates are required.   
We further evaluated the 2D performance with respect not only to the time step but also to the number of ``switching'' operations needed, finding that 2D performs well even under this metric (see Appendix.~\ref{sec:number_of_gates}-\ref{sec:time_step}).
We also compared 2D and 2T assuming that  the time-step resolution in 2T is not limited,  and found that the performance of 2D is still comparable to 2T even though this condition is in favor of 2T. We further find that 2D can outperform the  3rd-order Trotter expansion  and  the simplest sequence achieved by  alternating the two operators (see Appendix.~\ref{sec:Ruth}).

\paragraph*{Conclusions and outlook --}
Finding the exact global optimal Trotter sequence or even just the sequence minimizing  the minimal third-order error  can  be expensive to calculate. 
In this letter, we introduced  two Trotter sequences to achieve higher fidelity in digital quantum simulation, by exploiting a novel geometric framework to estimate the Trotterization error. 
By optimization the fidelity on a simple grid picture, our 2O and 2D sequences outperform the  widely used 2T sequence in complementary Trotter-step regimes. In addition, the 2D solution is based on an intuitive ordering that can be found with an efficient, greedy optimization algorithm.
Compared to 2T, less Trotter steps are required to reach the same fidelity with our sequences, thus also providing an advantage when the Trotter steps number is a performance metric. 
  
We note that while we mostly presented numerical results for two qubits, we numerically obtain similar performances  for larger systems, and indeed the scalings found are expected to apply to many-qubit systems. As the minimization algorithm is very efficient, since it does not require to evaluate quantum propagators, and it can be easily extended to deal with multiple building block operators, 2O and 2D can become  versatile and powerful tools to improve the fidelity of digital quantum simulation for many-qubit systems.
It could be interesting to further investigate applying our 2D and 2O sequences to other quantum algorithms beyond quantum simulation, such as quantum phase estimation~\cite{Abrams1999,Campbell19} or to find a good ansatz for the Quantum Approximate Optimization Algorithm~\cite{Farhi2014}.

\textbf{Acknowledgements}

\begin{CJK*}{UTF8}{}
 We thank Chao Yin \CJKfamily{gbsn}(尹超) for helpful discussions. This work was supported in part by DARPA DRINQS, NSF grant EECS1702716, and PHY1734011.
\end{CJK*}

\appendix

\section{Geometrical framework for Trotter error based on holonomy on a fiber bundle}
\label{sec:geometry}

In this section, we provide a geometrical perspective on our problem, based on the notion of holonomy on a fiber bundle~\cite{kobayashi}. 

Consider the space $G\times\R^2$ as a trivial bundle over $\R^2$, where $G=U(n)$ is the unitary group whose elements are unitary evolutions on the Hilbert space. We define a connection $\omega$ as follows:
\begin{equation}
\omega=Adx_1+Bdx_2,
\end{equation}
where $A$ and $B$ are $i\times$ the Hamiltonians in our problem.  This connection defines a parallel transport over the path $l$ as:
\begin{equation}
\mathcal{P}\exp(\int_l\omega),
\end{equation}
where $\mathcal{P}$ denotes path-ordering. 
Under this geometric setting, the error is nothing but the holonomy along a specific loop:
\begin{align}
&e^{-pA-qB}\left(e^{pa_1A/n}e^{qb_1B/n}\dots e^{pa_MA/n}e^{qb_MB/n} \right)^n \nonumber\\
&=\mathcal{P}\exp(\oint_l\omega),
\end{align}
where the loop $l$ goes from $(0,0)$  to $(p,q)$ along the path defining the Trotter sequence (as shown in Fig.~\ref{fig:geometric} of the main paper), and  then goes back to $(0,0)$ along the diagonal.

Geometrically, since the holonomy is provided by the curvature (the topology is trivial and therefore has no contribution), one expects that:
\begin{equation}\label{eq-expect}
\mathcal{P}\exp(\oint_l\omega)\approx\exp(\iint\Omega),
\end{equation}
where $\Omega$ is the curvature 2-form (with a good choice of gauge) and the double integral is over the area surrounded by the loop $l$. Indeed, it was proved in Ref.~\cite{expansion} that this expectation is true up to at least 3rd-order under a gauge in which the connection form is parallel in radial direction. {One can transform any connection $\omega$   into a  radially parallel form with a gauge transformation, resulting in:
}\begin{equation}
\frac{1}{2}\Omega_{ij}\int_l x_idx_j+\frac{1}{3}(\partial_k\Omega_{ij}+[\omega_k,\Omega_{ij}])\int_lx_ix_kdx_j+O(l^4).
\end{equation} 

In our case, 
\begin{equation}
\Omega=d\omega+\omega\wedge\omega=[A,B]dx_1dx_2,
\end{equation}
and a simple calculation gives (see Eqs.~\ref{eq:integral2}-\ref{eq:integral} of the main text) :
\begin{widetext}
\begin{equation}\label{eq-integral}
\begin{aligned}
&\frac{1}{2}[A,B](\int x_1dx_2-x_2dx_1)
+\frac{1}{3}[A,[A,B]]\int x_1(x_1dx_2-x_2dx_1)
+\frac{1}{3}[B,[A,B]]\int x_2(x_1dx_2-x_2dx_1)\\
=&[A,B]\iint dx_1dx_2+[A,[A,B]]\iint x_1dx_1dx_2
+[B,[A,B]]\iint x_2dx_1dx_2.
\end{aligned}
\end{equation}
\end{widetext}
Some nice features appear here:
\begin{itemize}
    \item This formula is true for arbitrary loops, not necessary piecewise linear as those in Fig.~\ref{fig:geometric} of the main paper.
    \item From the first line of Eq.~\eqref{eq-integral}, we see that the error can be expressed as an integral of \emph{local contributions} along the loop. This will make it possible for a dynamic programming algorithm to find the  global optimal path. 
    \item From the last line of Eq.~\eqref{eq-integral}), the coefficients of $[A,B]$ can be explained geometrically as the area surrounded by the loop, while the coefficients of $[A,[A,B]]$ and $[B,[A,B]]$ are moments about the $x_2$-axis and $x_1$-axis. Note that the areas and moments here are \emph{oriented}: positive for anti-clockwise boundary and negative for clockwise boundary; or equivalently (since the path can only go up or right), positive for areas on the right side of the diagonal and negative otherwise.
\end{itemize}

\section{Algorithm for constructing the 2-Optimal sequence}
\label{sec:DynamicProgramming}
In this section we provide one possible algorithm to find a path that has the minimal  third-order error (2O) on a 2-d $p\times q$ grid using dynamic programming. 
Using the first line of Eq.~\eqref{eq-integral} we can assign a triplet of weights  $\{\mathcal{E}_2,\mathcal{E}_{3A},\mathcal{E}_{3B}\}$ to each edge, where $\mathcal{E}_2$ is the second-order error and $\mathcal{E}_{3A},\mathcal{E}_{3B}$ are coefficients of $[A,[A,B]]$ and $[B,[A,B]]$ respectively in the third-order error. For convenience we associate two edges with node $(i,j)$ on the grid. $E_{r,i,j}$ is the vector weight for the edge connecting node $(i-1,j)$ and node $(i,j)$;  $E_{u,i,j}$ is the vector weight for  edge connecting node $(i,j-1)$ and node $(i,j)$. The algorithm has three steps:

\begin{enumerate}
\item Generate {\it{ErrorTable}}.  $ErrorTable(i,j)$ records the (cumulative) vector weight triplets of all possible paths from the origin --node $(0,0)$-- to node $(i,j)$. $ErrorTable$ can be generated recursively, See Algorithm~\ref{alg:step1}
\item Among the paths such that $ErrorTable(p,q)\{1\}=0$, find $MinErr=\min(|ErrorTable(p,q)\{2\}|+|ErrorTable(p,q)\{3\}|)$. 
\item Find the  path  that corresponds to $MinErr$ by tracing back from node $(p,q)$ to node $(0,0)$.
\end{enumerate}

Assuming $p\sim q\sim n$,  the total number of coefficients for the second-order term is $O(n^2)$, while the  number of coefficients for each third-order term  is $O(n^3)$. Therefore, the size of the error set at any point is at most $O(n^8)$. At each step in the loop, we need to calculate two additions, which cost $O(n^8)$ computational time, and a union, which also cost at least $O(n^8)$ time (with some good data structure like union-find or hash table). The total time needed is then $O(n^{10})$. 
We note that to find the optimal path, we need to optimize over the whole grid: 
While in the 2D finding algorithm we first normalize $p,q$ by their greatest common divisor, $g$, and repeat  $g$ times the optimal construction found for $(p/g,q/g)$, this simplification is not guaranteed to provide the correct solution in 2O. 
Indeed, one can verify that 2O does not possess translational symmetry; for example, the 2O ordering  for (4,4) is not simply repeating the 2O ordering (2,2) twice. Optimizing over all Trotter steps further increases the complexity.

\begin{algorithm}[H]
\caption{Construction of error table}
\label{alg:step1}
\begin{algorithmic} 
\STATE $E_{r,i,j} = (-j,  -j(i ^2 - (i - 1)^2), -2j^2)$
\STATE $E_{u,i,j} = (i, 2i^2, i (j ^2 - (j - 1)^2))$
\STATE $ErrorTable=Table[\{\},\{i,0,p\},\{j,0,q\}]$
\STATE $ErrorTable(0,0) = \{(0,0,0)\}$
\FOR{$1\leq i \leq p$ }
\STATE~~~~ $ErrorTable(i, 0) = ErrorTable(i - 1, 0) + E_{r,i,0}$
\ENDFOR
\FOR{$1 \leq j \leq q$}
\STATE~~~~ $ErrorTable(0, j) = ErrorTable(0, j-1) + E_{u,0,j}$
\ENDFOR
\FOR{$1\leq i \leq p$  }
\FOR{$1 \leq j \leq q$}
\STATE ~~~~$ErrorTable(i,j) =(ErrorTable(i-1,j) +E_{r,i,j})~ \bigcup ~(ErrorTable(i,j-1) +E_{u,i,j}) $
\ENDFOR
\ENDFOR
\end{algorithmic}
\end{algorithm}

\section{Proof that 2-Diagonal path is also the global minimal distance path }
\label{sec:D}

\begin{figure}[t]
  \centering
	\includegraphics[width=0.3\textwidth]{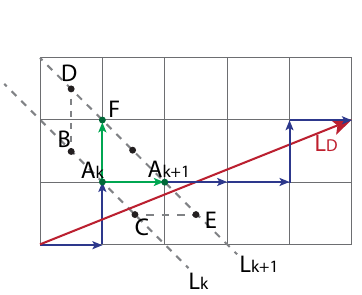}
  \caption{\textbf{Sketch of proving that 2D is also global optimal in terms of total distance.}}
\label{fig:proof}
\end{figure}

 In this section, we use mathematical induction to show that  any point $A_n$ on the 2D path $A_0 \rightarrow A_1\rightarrow\cdots\rightarrow A_{p+q}$ is the closest one to the diagonal among all possible points after $n$ steps.  So as a greedy method, 2D finds the global minimal distance path. 

To be more concrete, we want to show that for any $n$ and any $ x+y=n$, 
$dist(A_n,L_D)\leq dist((x,y),L_D)$, where $dist(\cdot,\cdot)$ is the distance between two inputs and and $L_D$ is the diagonal.

\begin{itemize}
\item When $n=0$, the statement is easily seen to be true.
\item Assume the induction hypothesis that 	$dist(A_k,L_D)\leq dist((x,y),L_D)$, where $x+y=k$, is true for any $k\geq0$.
\item From the induction hypothesis, we know that the diagonal $L_D$ intersects with $L_k: x+y=k$ between [B,C], where B and C are the centers of their unit grids, see Fig.~\ref{fig:proof}. We also know that $L_D$ points towards the upper right. So $L_D$ intersects with $L_{k+1}: x+y=k+1$ between (D,E), which guarantees that  at step $k+1$, either point F or $A_{k+1}$ is the closest to $L_D$. Between [D,E] there are two choices and we know that the hypothesis holds for k+1. 
\end{itemize}
 By mathematical induction, we know that the statement holds for any natural number n. Fig.~\ref{fig:proof} shows the least obvious case where the two possible steps following $A_k$ lie on the same side of $L_D$.

\section{Analytical bounds on 2-Diagonal fidelity}
\label{sec:appendix}
Assume that the target evolution is $e^{x(pA+qB)}$, where $p$ and $q$ are  mutually prime integers \footnote{If that is not the case and they have a common divider $m$, we can set  $p'=p/m$, $q'=q/m$ and find the construction for $U'=e^{x(p'A+q'B)}$ for $n$ and then repeat it $m$ times.}  and $x$ is a small parameters.

Approximating the desired evolution with a 2T expansion gives
\begin{align}
e^{\frac{1}{2}xpA}e^{xqB}e^{\frac{1}{2}xpA}	&= e^{x(pA+qB)} + x^3 \mathcal{E}  +O(x^4)
\label{eq:trotter1}
\end{align}
or 
\begin{align}
e^{\frac{1}{2}xqB}e^{xpA}e^{\frac{1}{2}xqB}	&= e^{x(pA+qB)} + x^3 \mathcal{E}'  +O(x^4)
\label{eq:trotterC2}
\end{align}
where $\mathcal{E}$ is the error operator. For Eq.~\eqref{eq:trotter1} we have
\begin{align}
	\mathcal E= -\frac{pq}{24}(pC+2qD),
\end{align}
and for Eq.~\eqref{eq:trotterC2} we have
\begin{align}
	\mathcal E'= -\frac{pq}{24}(2pC+qD),
\end{align}
where we defined
$C=[A,[A,B]]$ and $D=[B,[A,B]]$. 
The symmetrized 2D decomposition has a general  form 

\begin{align}
&e^{\frac{a_{M+1}}{2} xpA}e^{\frac{b_M}{2} xqB}\cdots e^{\frac{a_2}{2} xpA}(e^{\frac{b_1}{2} xqB}e^{a_1xpA}e^{\frac{b_1}{2} xqB})\nonumber \\	&e^{\frac{a_2}{2} xpA}\cdots e^{\frac{b_{M}}{2}xqB}e^{\frac{a_{M+1}}{2} xpA},
\label{eq:QInested}
\end{align}
where $\sum_{k=1}^{M+1} a_k= \sum_{k=1}^{M}b_i=1$, and $a_k\geq 0$, $b_k\geq 0$. 

We can evaluate the error by starting from the middle of the sequence and keeping track  of the third-order error:
\begin{align}
	e^{\frac{1}{2}b_1 xqB}e^{a_1xpA}e^{\frac{1}{2}b_1 xqB} &= e^{a_1 xpA +b_1 xqB} +x^3\mathcal{E}_1 +O(x^4)
\end{align}
The error $\mathcal{E}_1$ can be written as
\begin{align}
	\mathcal{E}_1&= 
\frac{pq}{24}a_1b_1(2p\,a_1C+q\,b_1D).
\end{align}
Consider the next layer
\begin{align}
	& e^{\frac{1 }{2}a_2 xpA} (e^{a_1 xpA +b_1 xqB} + x^3\mathcal{E}_1 )e^{\frac{1 }{2}a_2 xpA} \nonumber \\
	& = e^{(a_1+a_2) xpA +b_1 xqB} + x^3 (\mathcal{E}_2+\mathcal{E}_1) +O(x^4),
\end{align}
where 
\begin{align}
	\mathcal{E}_2&= 
-\frac{pq}{24}a_2b_1\left[p(a_2+2a_1)C+2q\,b_1D\right].
\end{align}
In general, we can find the formulas for the odd/even errors:
\begin{align}
	\mathcal{E}_{2k-1}&= 
\frac{pq}{24} b_k\mathcal A_k\left[2p\, \mathcal A_kC+q(2\mathcal B_{k-1}+b_k)D\right]~\text{and}\\
\mathcal{E}_{2k} &= -\frac{pq}{24} a_{k+1}\mathcal B_k\left[p(a_{k+1}+2\mathcal A_k)C+ 2q\,\mathcal B_kD\right],
\end{align}
where $\mathcal A_k=\sum_{i=1}^k a_i$ and $\mathcal B_k=\sum_{i=1}^kb_i$.
We want to compare the total error of 2D, $\mathcal E_{\textrm{2D}}=\sum_{k=1}^M (\mathcal E_{2k}+\mathcal E_{2k-1})$ with the conventional  2nd Trotter error $\mathcal E$ and $\mathcal E'$. We use the Frobenius norm to compare operators, $\|E\|^2_F=\textrm{Tr}(E^\dag E)$, and without loss of generality we assume $\|C\|_F=\|D\|_F$, and to simplify the presentation we also take $\textrm{Tr}(C^\dag D)=0$. If $\|C\|_F \neq\|D\|_F$ we can rescale $p$ to make $\|C\|_F=\|D\|_F$. Hereafter we can drop the subscript $F$ without causing confusion.   With these assumptions, it is straightforward to show that $\|\mathcal E_{\textrm{2D}}\|\leq \|\mathcal E\|$ iif $p^2\mathcal C^2+q^2\mathcal D^2\leq \text{min}(p^2+4q^2,q^2+4p^2)$, where 
\begin{align}
\mathcal C=&	\sum_{k=1}^M[ 2\mathcal A_k^2b_k-a_{k+1}\mathcal B_k(a_{k+1}+2\mathcal A_k)],
\\
\mathcal D=&	\sum_{k=1}^M[ \mathcal A_kb_k(b_k+2\mathcal B_{k-1})-2a_{k+1}\mathcal B_k^2].
\end{align}
Note that $\mathcal B_{k-1}$ only exists when $k\geq 2$.  Indeed, we have 
\begin{equation}\label{eq:error}
\mathcal E_{\textrm{2D}}=\frac{pq}{24}(p\,\mathcal C\, C+ q\,\mathcal D\, D)\end{equation}

If we rewrite Eq.~\eqref{eq:error} in form of integration, Eq.~\eqref{eq-integral} is regained. 
 Define a function $f(a_1,\cdots,a_{M+1},b_1,\cdots,b_{M})=p^2\,\mathcal C^2+q^2\,\mathcal D^2$. We can  show that any choice of the coefficients $a_i$, $b_i$ that does not collapses the 2D expansion onto the two 2nd Trotter ones gives a lower error than $\text{max}(\|\mathcal{E}\|, \|\mathcal{E}'\|)$. The strategy is to find the upper bound of $f$ by varying $a_i$ and $b_i$ recursively. 

 We can start by varying the $a_{M+1}$ parameter. First change $a_1,a_2,\cdots,a_M,a_{M+1}$ to a set of independent parameters $\mathcal{A}_{1}, \mathcal{A}_2,\cdots, \mathcal{A}_{M-1},a_M=1-a_{M+1}-\mathcal{A}_{M-1},a_{M+1}$. After this change $a_{M+1}$ is an independent variable.  All the terms in $\mathcal C$ that depend on $a_{M+1}$ are $3b_M(1-a_{M+1})^2-\mathcal B_M+\mathcal A_{M-1}(\mathcal B_M-b_M)$ therefore
 
\begin{align}
	\frac{\partial \mathcal C}{\partial a_{M+1}} &=-6b_M(1-a_{M+1})\leq 0, ~\text{and} \label{eq:1} \\
	\frac{\partial^2 \mathcal C}{\partial a_{M+1}^2} &=6b_M \geq 0. \label{eq:2}
\end{align}
$\mathcal C(a_{M+1})$ is a decreasing function and $\mathcal C (a_{M+1}=0)=2$, $\mathcal C(a_{M+1}=1)=-1$. Actually $ C(a_{M+1}) =3b_M a_{M+1}^2-6b_M a_{M+1}+2$.

Similarly we find 
\begin{align}
	\frac{\partial \mathcal D}{\partial a_{M+1}} &= 3b_M(b_M-2\mathcal B_M)\leq 0 , ~\text{and} \label{eq:3}\\
	\frac{\partial^2 \mathcal D}{\partial a_{M+1}^2} &=0. \label{eq:4}
\end{align}
$\mathcal D(a_{M+1})$ is a decreasing function and $\mathcal D (a_{M+1}=0)=1$, $\mathcal C(a_{M+1}=1)=-2$. Actually $\mathcal D(a_{M+1})= 3b_M(b_M-2)a_{M+1}+1$.

The derivatives of the function $f(a_1,\cdots,a_{M+1},b_i,\cdots,b_{M})=p^2\,\mathcal C^2+q^2\,\mathcal D^2$ are given by 
\begin{align}
\frac{\partial f}{\partial a_{M+1}} &=2p^2\mathcal C \frac{\partial \mathcal C}{\partial a_{M+1}} + 2q^2 \mathcal D	\frac{\partial \mathcal D}{\partial a_{M+1}},~\text{and}\\
\frac{\partial^2 f}{\partial a_{M+1}^2} &=  2p^2 (\frac{\partial \mathcal C}{\partial a_{M+1}})^2 +2p^2\mathcal C\frac{\partial^2 \mathcal C}{\partial a_{M+1}^2} \nonumber \\
&+ 2q^2 (\frac{\partial \mathcal D}{\partial a_{M+1}})^2 +2q^2\mathcal D\frac{\partial^2 \mathcal D}{\partial a_{M+1}^2}. \label{eq:5}
\end{align}
Plugging  Eq.~\eqref{eq:1},\eqref{eq:2},\eqref{eq:3},\eqref{eq:4} into \eqref{eq:5} we find that when $\mathcal C >0$, $f$ is a convex function of $a_{M+1}$ and we have $f\leq \text{max}\{f(a_{M+1}\!=\!0),  f(a_{M+1}\!=\!1)\}$. When $\mathcal C \leq 0$, it is not difficult to verify that when $\mathcal C \leq 0$, we also have $\mathcal D \leq 0$. Even if $\frac{\partial^2 f}{\partial a_{M+1}^2}\leq 0$, we still have $\frac{\partial f}{\partial a_{M+1}} \geq 0$ and $f\leq \text{max}\{f(a_{M+1}\!=\!0),  f(a_{M+1}\!=\!1)\}$ is still true. 

When $a_{M+1}=1$ (that is, all other coefficients $a_1,\cdots,a_{M}$ are zero, meaning that $M=0$) $\mathcal C=-1$, $\mathcal D=-2$ and $f=p^2+4q^2$, as expected, since we retrieve the simpler Trotter expansion $e^{\frac{1}{2}xpA}e^{xqB}e^{\frac{1}{2}xpA}$. 

When $a_{M+1}=0$, we can repeat the same analysis above, now taking the derivatives with respect to $b_M$. We find again that for $b_M=1$ the construction collapses to the Trotter expansion $e^{\frac{1}{2}xqB}e^{xpA}e^{\frac{1}{2}xqB}$, and we can analyze the case $b_M=0$ as done above. Recursively, we thus find that  $f\leq 4p^2+q^2$, with equality in the case where the construction simplifies to the Trotter expansion $e^{\frac{1}{2}xqB}e^{xpA}e^{\frac{1}{2}xqB}$ or $f\leq p^2+4q^2$, with equality in the case where the construction simplifies to the Trotter expansion $e^{\frac{1}{2}xpA}e^{xqB}e^{\frac{1}{2}xpA}$.

When $p=q$, this proves that  any decomposition of the exponential operators with more and smaller steps than 2T yields a smaller error than 2T. 

When $p\neq q$,  the construction following the form in Eq.~\eqref{eq:QInested} has an error that is bounded by $\text{max}(\|\mathcal{E}\|, \|\mathcal{E}'\|)$. We can further show that the error of 2D is actually bounded by $\text{min}(\|\mathcal{E}\|, \|\mathcal{E}'\|)$. 
When $p>q$, $\|\mathcal{E}\|<\|\mathcal{E}'\|$ and we want to show that  $f\leq p^2+4q^2$.  According to the construction of 2D, when $p>q$, we first take $m$ steps along $A$, where $\frac{mq}{p+q}<\frac{1}{2}$ and $\frac{(m+1)q}{p+q}> \frac{1}{2}$, i.e., $m=\lfloor\frac{p+q}{2q}\rfloor$. The construction that is closest to $e^{\frac{q}{2}xB}e^{xpA}e^{\frac{q}{2}xB}$ allowed by 2D writes
\begin{equation}
e^{\frac{2m}{p}\frac{p}{2}xA}e^{\frac{q}{2}xB}e^{(1-\frac{2m}{p})pxA}e^{\frac{q}{2}xB}	e^{\frac{2m}{p}\frac{p}{2}xA}.
\label{eq:5piece}
\end{equation}
For the decomposition in  Eq.~\eqref{eq:5piece}, we find  that $\mathcal{C} = 3 (1-\frac{2m}{p})^2-1$ and $\mathcal{D }  = 1-\frac{6m}{p}$. It's not difficult to verify that  $p^2\mathcal{C}^2+q^2\mathcal{D}<p^2+4q^2$, when $p>q$. If we further decompose the central part $e^{\frac{q}{2}B}e^{(1-\frac{2m}{p})pA}e^{\frac{q}{2}B}$, since $q\geq 1-\frac{2m}{p}$, the worst case will converge to $e^{(\frac{1}{2}-\frac{m}{p})pA}e^{qB}e^{(\frac{1}{2}-\frac{m}{p})pA}$, with which the whole formula becomes $e^{\frac{1}{2}pA}e^{qB}e^{\frac{1}{2}pA}$. In other words, however you decompose the central part in Eq.~\eqref{eq:5piece}, the error of the whole formula is bounded by $p^2+4q^2$. 
In summary, when $p>q$, if we  construct the nested structure according to the principle of 2D, the error will be smaller than $p^2+4q^2$. Similarly when $p<q$, $f \leq q^2+4p^2$. In summary  $f \leq \text{min} (p^2+4q^2,q^2+4p^2)$ and thus $\|\mathcal{E}_{2D}\|\leq \text{min}(\|\mathcal{E}\|,\|\mathcal{E}'\|)$.

\section{Extension to three operators}
\label{sec:3D}
Using the geometric picture, 2D can be easily extended to three or more operators. The greedy algorithm always find the step that keeps the trajectory closest to the diagonal. {In principle one could also use dynamic programming to find a path that minimizes the third-order error. Nevertheless, as the number of non-commutative operators increases, the complexity of the algorithm increases fast, while 2D remains efficient. }
Here we tested the performance of 2D on a two-qubit Ising Hamiltonian with both transverse and longitudinal fields.  The three implementable Hamiltonians read
\begin{equation*}
H_1 = \sigma_z^1  \sigma_z^2, \qquad
H_2 = \frac{1}{2}(\sigma_z^1 + \sigma_z^2),\qquad
H_3 = \frac{1}{2}(\sigma_x^1 + \sigma_x^2)
\end{equation*}
and the  target evolution is $e^{-i(pH_1 +qH_2 +r H_3)t}$, where $p, q, r$ are integers with greatest common divisor 1.  

\begin{figure}[ht]
  \centering
	\includegraphics[width=0.5\textwidth]{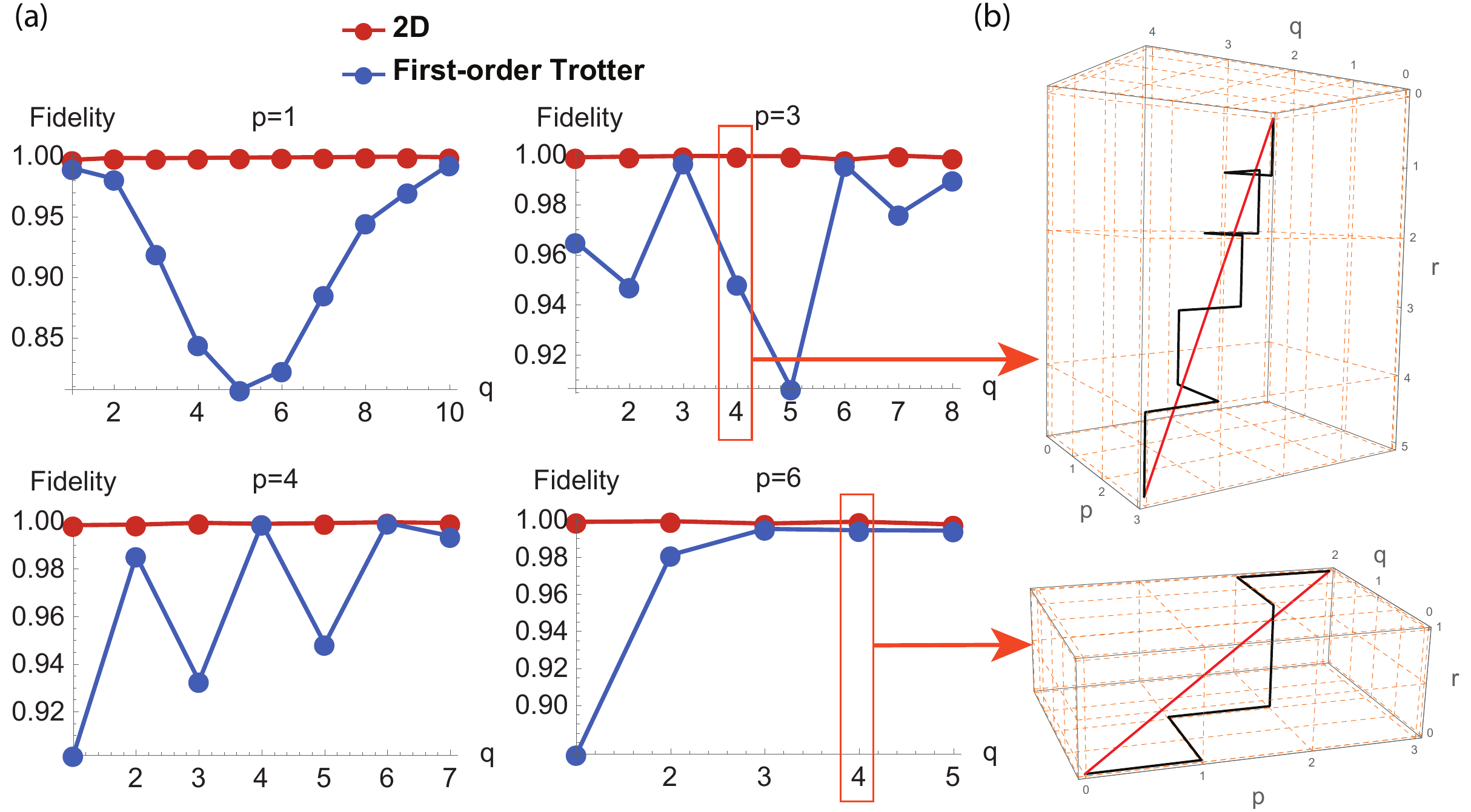}
  \caption{\textbf{2-Diagonal for three operators.} {(a)} $t=1$. $n=10$. $p+q+r=12$. {(b)}The trajectory of 2D for two examples (3,4,5) and (6,4,2). }
\label{fig:3D}
\end{figure}

We compare  in Fig.~\ref{fig:3D}(a) the results for 2D and the first-order Trotter expansion (for simplicity, we did not symmetrize either 2D or first-order Trotter, so that both are now first-order approximations). For case $p=3,~q=4,~r=5$, first-order Trotter expansion is $(e^{-i 3H_1 \frac{t}{n} }e^{-i 4 H_2 \frac{t}{n}}e^{-i 5 H_3 \frac{t}{n}})^{n}$ and 2D is 
\begin{align*}
&(e^{-i H_3\frac{t}{n}}e^{-i H_2\frac{t}{n}}e^{-i H_1\frac{t}{n}}e^{-i H_3\frac{t}{n}}e^{-i H_2\frac{t}{n}}e^{-i H_1\frac{t}{n}}\\
\cdot &e^{-i H_3\frac{t}{n}}e^{-i H_2\frac{t}{n}}e^{-i H_3\frac{t}{n}}e^{-i H_1\frac{t}{n}}e^{-i H_2\frac{t}{n}}e^{-i H_3\frac{t}{n}})^n. 
\end{align*}
For case $p=6,~q=4,~r=2$, first-order Trotter expansion is $(e^{-i 3H_1 \frac{t}{n} }e^{-i 2 H_2 \frac{t}{n}}e^{-i 1 H_3 \frac{t}{n}})^{2n}$ and 2D expansion is 
\begin{align*}
&(e^{-i H_1\frac{t}{n}}e^{-i H_2\frac{t}{n}}e^{-i H_1\frac{t}{n}}e^{-i H_3\frac{t}{n}}e^{-i H_2\frac{t}{n}}e^{-i H_1\frac{t}{n}})^{2n}.
\end{align*}
The trajectories of the two cases are shown in Fig.~\ref{fig:3D}(b). We see that 2D outperforms the first-order Trotter expansion for all $(p,q,r)$s up to $p+q+r=12$. Similar results can be obtained with different Hamiltonians.  

\section{Number of gates and number of Trotter steps}

\label{sec:number_of_gates}
\begin{figure*}
  \centering
	{\includegraphics[width=\textwidth]{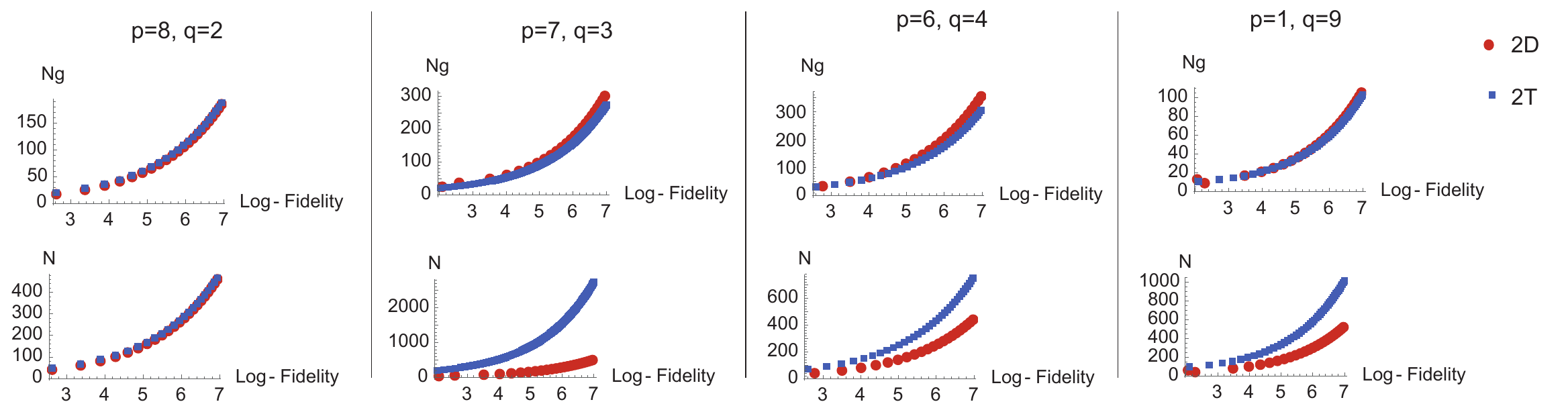}}
  \caption{\textbf{Number of gates (switchings) and number of Trotter steps.} $t=1$, $p+q=10$. Upper panel: total number of gates (switchings) needed to reach a certain fidelity for 2D (red) and 2T (blue). Lower panel: total number of Trotter steps. For $p=8,~q=2$, 2D and 2T has the same ordering and they have the same performance. Here $A=- i H_1 t/n$, $B= - i H_2 t/n$, where $H_1 = \frac{1}{2}(\sigma_z^1 + \sigma_z^2$), and $H_2 = \sigma_x^1 \sigma_x^2$. }
\label{fig:number_of_operations}
\end{figure*}

While 2D reaches better fidelity than 2T, it generally does so by increasing the number of alternations between the two (or more) building blocks $e^A$ and $e^B$. 
When simulating spin dynamics, the total number of switches and gates is not so crucial since the number of Hamiltonians is not large, and the total simulation time remains the same for different Trotter expansions.
In many experimental implementations, switching between the two Hamiltonians comes at a cost, either because of additional time overheads, or because the switch itself, obtained, e.g., by applying a short control pulse, might introduce control errors.  Still, large number of pulses have routinely been applied, e.g., in dynamical decoupling~\cite{Abobeih18,Bar-Gill13} and  Hamiltonian engineering~\cite{Wei18}, so a large number of switchings is not infeasible.  Quantifying the decrease in fidelity due to such errors is complex, because it would be platform dependent. Still, we can evaluate the number of switchings required to achieve a certain fidelity for different constructions.  We find that 2D does not always impose an overhead in the number of switches with respect to, e.g., the 2T expansion, although this does depend on the particular $(p,q)$ pair considered (see Fig.~\ref{fig:number_of_operations}). In particular, for more ``unbalanced'' pairs ($p\ll q$ or vice-versa), 2D can outperform 2T.

When the  elementary unitary evolutions are composed by a set of discrete gates that cannot be merged~\cite{Jones_2012}, the number of switches no longer quantify the amount of resources needed. The total number of Trotter steps, which is proportional $(p+q)n$, best describes the resources needed. The lower panels in Fig.~\ref{fig:number_of_operations} shows that to reach the  same fidelity, 2D requires a smaller number of Trotter steps, thus providing an advantage with respect to this metric. 

\section{Performance when the time-step resolution is not limited}
\label{sec:time_step}

\begin{figure}[ht]
  \centering
	\includegraphics[width=0.35\textwidth]{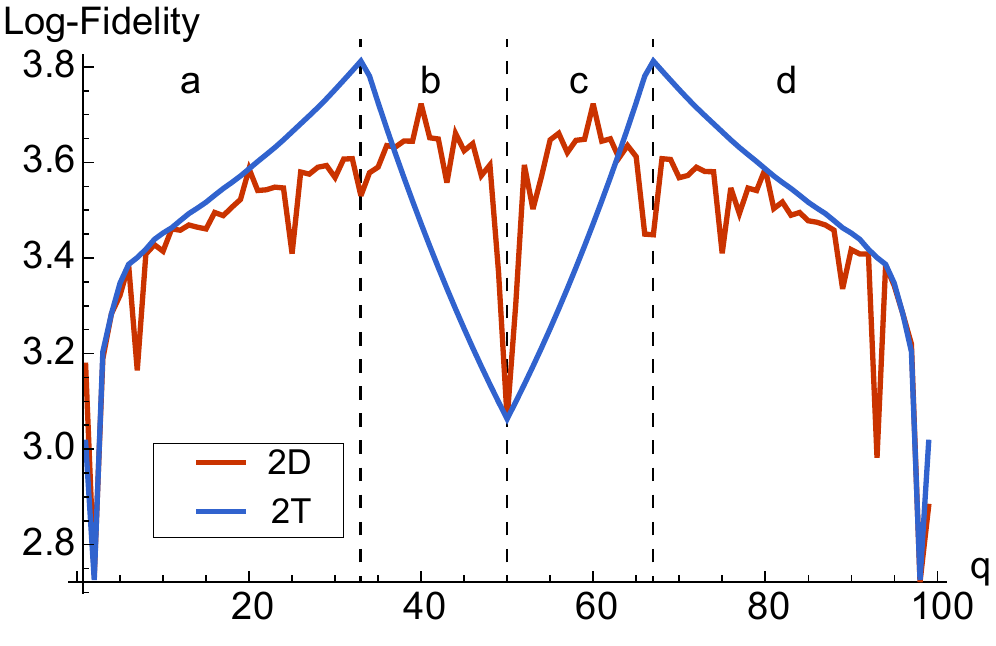}
  \caption{\textbf{Comparison between 2D and 2T with infinitely high time resolution.} The four region a, b, c, and d correspond to $p\geq 2q$, $q \leq p<2q$, $q/2\leq p<q$, and $p< q/2$ respectively, where the best Trotter expansion differ. In this calculation,  $H_1 = \frac{1}{2}(\sigma_z^1 + \sigma_z^2$) and $H_2 = \sigma_x^1 \sigma_x^2$ are used. $p+q$ is fixed at 100, $t=1$ and  $n=20$. We choose a number of Trotter steps, $n=20$, that is in the range of what used in digital quantum simulation with super conducting qubits~\cite{Salathe15} (where $J\sim$40MHz, $\Delta t \sim 1ns$ and $1/n=J\Delta t=1/25$) and in the manipulation of relatively strongly coupled nuclear spins with Nitrogen Vacancy center in diamond~\cite{Abobeih18}.}
\label{fig:6}
\end{figure}

2D outperforms 2T under the constraints that only integer number of elementary gates are implementable.  For example $e^{A/n}$ is implementable but $e^{1.2A/n}$ is not allowed. The constraints could come from experimental limitations on both the shortest time step and the time step resolution.  Often in practical implementations, the duration of each Trotter step, even when it  is above the smallest time step, cannot be set at will,  but it is typically an integer multiple of a minimum time duration. We have assumed that this minimum time duration is the smallest Trotter step.  Qubits with the fastest dynamics suffer from these two limitations since typically the timing of each operation is set by a digital clock with a minimum resolution. In addition, often, a single Trotter step is given by a fixed (fault-tolerant) quantum gate (such as CNOT gate), thus imposing an effective minimum ``time resolution'' constraint. 
However, for  qubits that present a slow dynamics, the time-resolution constraint might be negligible. In this case, assuming that there are no constraints on time resolution, 2T typically performs better than  2D that by construction assumes a finite time-step resolution (equal to the minimum time step).  In Fig.~\ref{fig:6} we compare 2T and 2D constructions for different $p,~q$. Note that for different $p$, $q$ values, there is an optimal 2T expansion, which is the one we selected. Still, we see that 2D has better performance than 2T for certain $p$, $q$ values.

\section{Comparison with other Trotter expansions}
\label{sec:Ruth}
\begin{figure*}
  \centering
	{\includegraphics[width=0.9\textwidth]{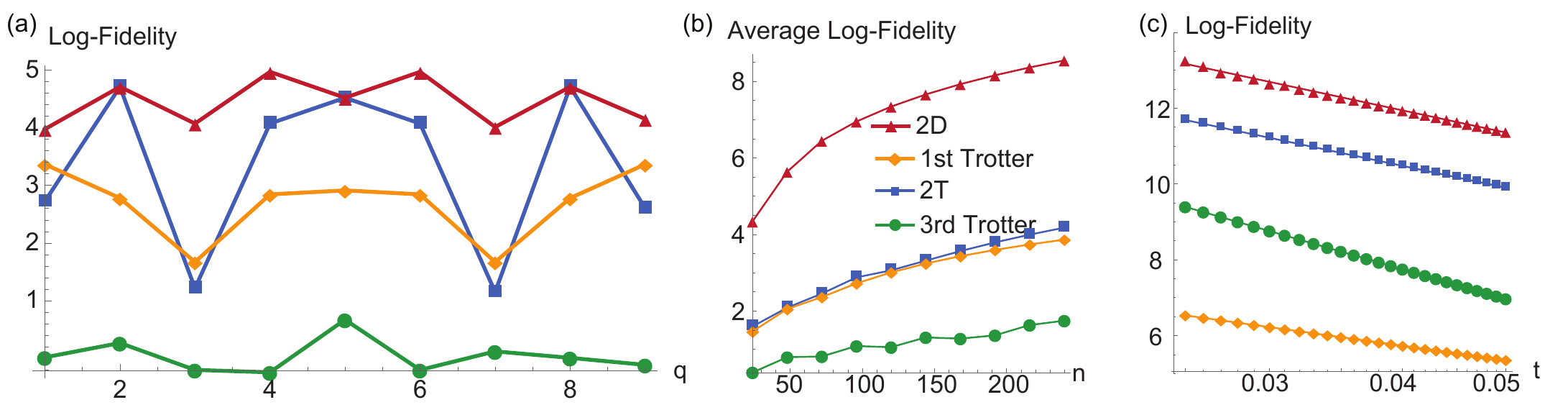}}
  \caption{\textbf{Fidelity comparison.} (a) Log-fidelity over the full range of possible engineered Hamiltonians, with $p+q=10$, $t=1$, $n=10$.  Here $A=- i H_1 t/n$, $B= - i H_2 t/n$, where $H_1 = \frac{1}{2}(\sigma_z^1 + \sigma_z^2$), $H_2 = \sigma_x^1 \sigma_x^2$. The total number of steps is similar to the total number of Floquet cycles in circuit cQED experiments~\cite{Langford17}. Similar parameters have also been used in NMR experiments~\cite{Alvarez15}, nuclear spins associated with the NV center in diamond~\cite{Abobeih18}, and ion traps~\cite{Zhang17n}. Log-fidelity is defined as $-\log_{10}(1-F)$, where $F$ is defined in Eq.~(\ref{eq:fidelity}) in main paper.
(b) Log-fidelity averaged over all possible $p+q=100$ as a function of $n$.  Note that the third-order does not perform as well, since the complex coefficients in Ruth's formula impose larger time steps. (c) Log-fidelity as a function of $t$. $p=6$, $q=7$, $n=24$. Solid lines are  fittings to $-\log_{10}(1-F) = a \log_{10}t +b$. Slopes of 2D, 2T, first-order, and third-order Trotter (Ruth's formula) are -6.3, -5.9, -3.9, and -8.1 respectively. From Taylor expansion the slopes should be -6, -6, -4, and -8.  Because in this plot $t$ is smaller than in (a) \&(b), the better scaling with $t$ of the third-order expansion prevails, yielding better fidelity than the first-order Trotter.}
\label{fig:fidelity}
\end{figure*}

Higher order Trotter expansions have the general form of $e^{p A + qB}\approx e^{c_1 pA}e^{c_2 qB}e^{c_3 pA}\cdots e^{c_M qB}$, where however the coefficients $pc_{1,3,\cdots}$ and $qc_{2,4,\cdots}$  are usually not integers and can even be negative and thus typically challenging (or impossible) to implement experimentally.  One example of third order Trotter expansion is Ruth's formula~\cite{Ruth83}, 
\begin{align}
		 (e^{{7\over24n}pA}e^{{2\over3n}qB}e^{{3\over4n}pA}e^{-{2\over3n}qB}e^{-{1\over24n}pA}e^{{qB}})^n&\approx e^{p A + qB},
\label{eq:ruth}
\end{align}
where negative coefficients are used. Indeed, some negative coefficients are unavoidable in third- and higher-order Trotter expansions~\cite{Suzuki91}. Implementing ``time reversal" to construct the unitaries with negative coefficients is not always possible. In Fig.~\ref{fig:fidelity} we show that when Trotter step is not very small ($t/n=0.1$), 2nd-Near-optimal has better fidelity than the conventional first-, second-, and third-order Trotter expansions. In the numerical evaluation, we in total $(p+q)n=100$ Trotter steps are used. For 100 steps it is difficult to find 2O path so we only focus on 2D.

\begin{figure*}[t]
  \centering
	{\includegraphics[width=0.99\textwidth]{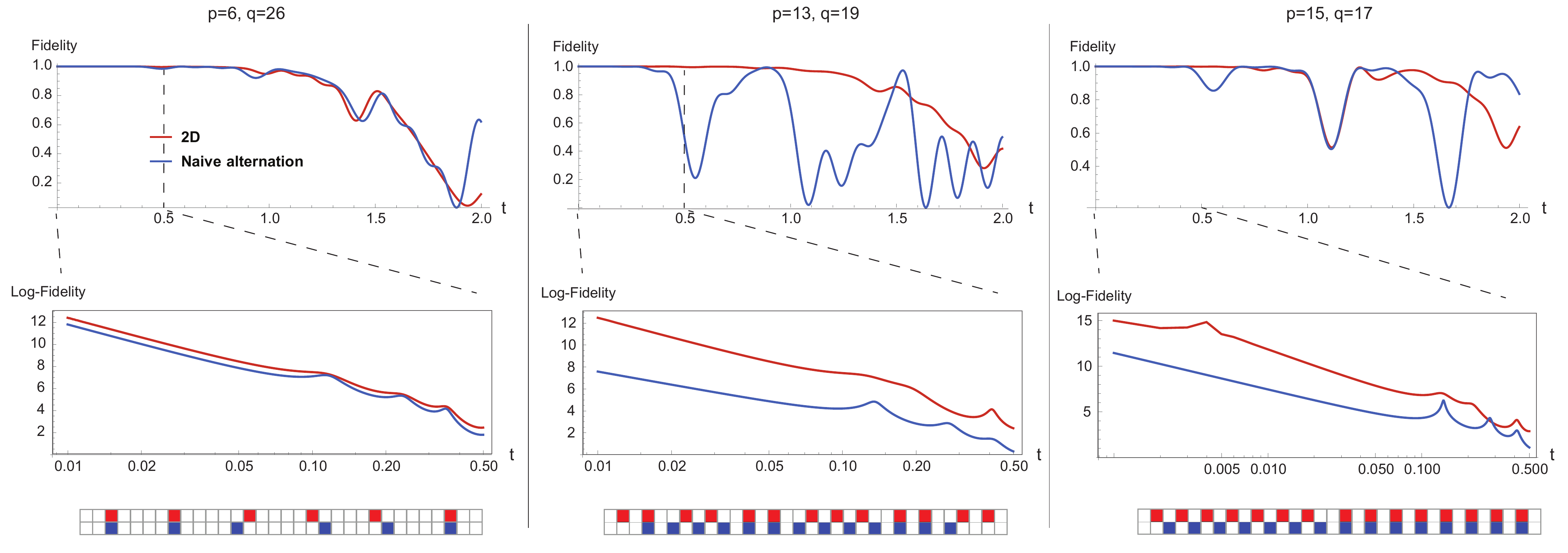}}
  \caption{\textbf{Comparison with naive alternation.}  
We compare the fidelity of 2D with constructions naively obtained by alternating as much as possible the two basic operators, $e^A$ and $e^B$. The fidelity as a function of the evolution time $t$ is better for 2D than for the naive constructions. We consider three cases, all with $p+q=32$ and $p=6$, $13$ and $15$, and we set $n=4$. Log fidelity in small $t$ regime is  plotted in log scale to better show the difference between different constructions (Lower panels). The 2D and naive constructions are pictorially depicted with white boxes indicating $e^B$ and red (blue) boxes $e^A$ for the 2D (naive) construction.}
\label{fig:naive_alternation}
\end{figure*}

While the comparison to well-known Trotter decomposition is instructive, one might think that simple alternative constructions could be found heuristically. For example, the intuition behind 2D is to maximize the alternation (spreading out) of the two operators,   
$e^{A/n}$ and $e^{B/n}$. For many $(p,q)$ pairs, however, it is not simple to find an exact alternating ordering. Then, a naive alternation construction is still outperformed by 2D, see, e.g., Fig.~\ref{fig:naive_alternation} for several examples. 

\bibliographystyle{apsrev4-1}
\bibliography{QI}

\end{document}